\documentclass[sigconf,nonacm]{acmart}

\usepackage{multirow}
\usepackage{multicol}
\usepackage{rotating}
\usepackage{hyperref}
\usepackage{siunitx}
\usepackage{stfloats}

\newcommand{\No}{\textit{8M$\_$easy}}
\newcommand{\Medium}{\textit{8M$\_$difficult}}
\newcommand{\High}{\textit{8M$\_$impossible}}
\newcommand{\Cmixed}{\textit{0M$\_$mixed}}
\newcommand{\Cclear}{\textit{0M$\_$clear}}

\usepackage{array,booktabs}
\newcommand {\otoprule}{\midrule [\heavyrulewidth]}
\newcolumntype {+}{ >{\global\let\currentrowstyle\relax}}
\newcolumntype {^}{ >{\currentrowstyle }}
  \newcommand {\rowstyle}[1]{\gdef\currentrowstyle{#1} %
  #1\ignorespaces
  }
\newcommand{\tabhead}{\rowstyle{\bfseries}}

\AtBeginDocument{%
  \providecommand\BibTeX{{%
    \normalfont B\kern-0.5em{\scshape i\kern-0.25em b}\kern-0.8em\TeX}}}

\copyrightyear{2022} 
\acmYear{2022} 
\setcopyright{rightsretained} 
\acmConference[CHI '22]{CHI Conference on Human Factors in Computing Systems}{April 29-May 5, 2022}{New Orleans, LA, USA}
\acmBooktitle{CHI Conference on Human Factors in Computing Systems (CHI '22), April 29-May 5, 2022, New Orleans, LA, USA}
\acmDOI{10.1145/3491102.3501915}
\acmISBN{978-1-4503-9157-3/22/04}

\begin{document}

\title{How Accurate Does It Feel? -- Human Perception of Different Types of Classification Mistakes}

\author{Andrea Papenmeier}
\email{andrea.papenmeier@gesis.org}
\affiliation{%
  \institution{GESIS -- Leibniz Institute for the Social Sciences}
  \streetaddress{Unter Sachsenhausen 6-8}
  \city{Cologne}
  \country{Germany}
  \postcode{50667}
}
\author{Dagmar Kern}
\email{dagmar.kern@gesis.org}
\affiliation{%
  \institution{GESIS -- Leibniz Institute for the Social Sciences}
  \streetaddress{Unter Sachsenhausen 6-8}
  \city{Cologne}
  \country{Germany}
  \postcode{50667}
}
\author{Daniel Hienert}
\email{daniel.hienert@gesis.org}
\affiliation{%
  \institution{GESIS -- Leibniz Institute for the Social Sciences}
  \streetaddress{Unter Sachsenhausen 6-8}
  \city{Cologne}
  \country{Germany}
  \postcode{50667}
}
\author{Yvonne Kammerer}
\email{kammerer@hdm-stuttgart.de}
\affiliation{%
  \institution{Stuttgart Media University}
  \streetaddress{Nobelstraße 10}
  \city{Stuttgart}
  \country{Germany}
  \postcode{70569}
}
\author{Christin Seifert}
\email{christin.seifert@uni-due.de}
\affiliation{%
  \institution{University of Duisburg-Essen}
  \streetaddress{Hufelandstraße 55}
  \city{Essen}
  \country{Germany}
  \postcode{45122}
}

\renewcommand{\shortauthors}{Papenmeier et al.}

\begin{abstract}
  Supervised machine learning utilizes large datasets, often with ground truth labels annotated by humans. While some data points are easy to classify, others are hard to classify, which reduces the inter-annotator agreement. This causes noise for the classifier and might affect the user's perception of the classifier's performance.  In our research, we investigated whether the classification difficulty of a data point influences how strongly a prediction mistake reduces the ``perceived accuracy''. In an experimental online study, 225 participants interacted with three fictive classifiers with equal accuracy (73\%). The classifiers made prediction mistakes on three different types of data points (easy, difficult, impossible). After the interaction, participants judged the classifier's accuracy. We found that not all prediction mistakes reduced the perceived accuracy equally. Furthermore, the perceived accuracy differed significantly from the calculated accuracy. To conclude, accuracy and related measures seem unsuitable to represent how users perceive the performance of classifiers.
\end{abstract}




\maketitle

\section{Introduction}

Artificial intelligence (AI), and particularly machine learning-based methods, are key to many data-driven applications in various fields, e.g., financial risk assessment~\cite{emerson2019trends}, hiring~\cite{Liem2018}, medical diagnosis~\cite{esteva2017dermatologist}, music recommendations~\cite{Wang2014}, or social media~\cite{Alvarado2018}. Throughout the years, machine learning systems have been optimized to increase their performance and produce better results. So far, however, the evaluation has mainly focused on the system perspective: Machine learning systems are often trained and evaluated on data that contains noise. But what if the quality of the underlying data is not as good as expected and does not reflect the real world? That would result in machine learning systems that are nearly perfect but might be perceived as performing poorly by humans. Consequences might be mistrust in the system~\cite{Dzindolet2002, dietvorst2015algorithm, Dietvorst2018}, a bad user experience~\cite{Fairclough2015accuracy}, or, even worse, severe consequences that affect people's life, e.g., introducing or replicating racist or cultural bias~\cite{Mehrabi2021, O'Neil2017, Noble2018}.


With the upcoming interest in a more socio-technical view on AI (e.g., \cite{Amershi2014, Bansal2019, shneiderman2020human, wortman2021human}), data work is recently gaining attention. Undervalued data quality causes cascading events that -- often unnoticeably -- impair the performance of machine learning systems ~\cite{Sambasivan2021}. Given a dataset including ground truth labels, supervised machine learning systems learn the underlying function that maps data to labels. Those labels are often retrospectively added to the data by human annotators. However, this might be challenging due to classification difficulties, such as ambiguity in the data~\cite{Tsipras2020contextualizing} or disagreement amongst annotators. A low inter-annotator agreement~\cite{Dumitrache2018} causes label noise in the data~\cite{algan2021image, Pavlick2019, Sheng2008}. Furthermore, even experts cannot always annotate data free of doubt~\cite{algan2021image} and might create a controversial instead of a real ground truth. Researchers address label noise by considering, for example, individual annotator votes for a data label instead of a majority vote~\cite{Resnick2021}. 

However, in some cases, several different labels are quite plausible, and prediction mistakes (according to the ground truth based on one of these labels) might not be perceived as mistakes by humans~\cite{Tsipras2020contextualizing}. For example, a human annotator might label an image as ``monastery'', while a machine learning system predicts the label ``church''~\cite{Tsipras2020contextualizing}. A user would probably accept both labels. Similarly, the news headline ``Keeping A Clear Mindset During School'' (taken from the News Category Dataset~\cite{dataset}) was labeled with the topic ``college'', although ``education'' could also be an acceptable label. Ultimately, this means that system performance does not always correspond to how users perceive a system: It can be better than assumed~\cite{Tsipras2020contextualizing} but also worse~\cite{Gordon2021}.

In this paper, we follow the line of research on user perception of the performance of machine learning systems~\cite{Gordon2021, Kay2015how, Kocielnik2019accept, Tsipras2020contextualizing} and investigate to what extent different levels of classification difficulty of single data points affect how users perceive the performance of a classifier. To analyze the effects of classification difficulty, we first set up a crowdsourcing task (N = 54)  to identify data points that are easy-to-classify, difficult-to-classify, and impossible-to-classify for humans for a topic classification dataset. We used a dataset that contains ground truth labels for user-formulated requirements on a new laptop and jacket. For example, \textit{``it needs to have lots of pockets and a hood too''} is easy to classify, while \textit{``I am going to buy a new cheap one''} is an impossible-to-classify sentence for which both labels are plausible. We then created two subsets for experimentation: One mixed dataset containing an equal number of easy-to-classify, difficult-to-classify, and impossible-to-classify data points and a clear dataset with only easy-to-classify data points. Subsequently, we manually produced the output of three fictive classifiers that made topic predictions for each data point with an accuracy of 73\% and two fictive classifiers with a calculated accuracy of 100\%. In an experimental online study (N = 225), we confronted participants with the predictions of one out of the five classifiers and asked them to assess the classifier's accuracy -- we call this the ``perceived accuracy''.

Our findings revealed a significant difference in perceived accuracy between all three classifiers with 73\% calculated accuracy: The more difficult it is for humans to correctly classify a data point, the less harmful is a prediction mistake on that data point.
As a result, traditional performance measures (e.g., accuracy, precision, recall, F1, ROC AUC) do not accurately reflect the level of perceived accuracy. We further found that the dataset composition plays an important role: A perfect classifier of the mixed dataset was perceived as significantly less accurate than a perfect classifier on the clear dataset, although both had a calculated accuracy of 100\%. These findings have important implications for how we should evaluate machine learning systems. Instead of focusing solely on accuracy, we suggest including user-centered factors such as classification difficulty in evaluation metrics.

\section{Related Work}
\label{sec:related_work}
In the following section, we contextualize our research in related work. First, we provide a short introduction to performance measures for machine learning systems that we will use to compare our results with. Then, we illustrate a common technical approach to deal with different prediction mistakes. How ground truth labels are gathered and how they contribute to noise in machine learning systems is discussed in \ref{ssec:rel_truth}. The following subsection shows how humans perceived systems’ accuracies and prediction mistakes. The related work section concludes with an overview of how machine learning systems can be evaluated from a more human-centered perspective.  

\subsection{Performance of Machine Learning Systems}
\label{ssec:rel_performance}
The performance of classifiers in supervised machine learning can be assessed with different measures, such as accuracy, sensitivity,
\vfill\eject\noindent
specificity, precision, recall, F1-score, ROC-curve, and ROC AUC~\cite{Bradley1997AUC,sokolova2006beyond}. In binary classification, those eight measures can be defined based on the four possible outcomes of a two-class problem\footnote{Here we use the common class labels ``positive'' and ``negative'', originating from information retrieval where an example can be relevant or irrelevant, i.e., ``positive'' or ``negative''. The definition of measures extends to any two-class problem, regardless of the class label names.}: true positive predictions ($tp$) and true negative predictions ($tn$), i.e., examples correctly classified as positive or negative, and their incorrectly classified counterparts false positives ($fp$) and false negatives ($fn$). Accuracy is defined as the ratio of correct predictions ($tp+tn$) and total number of predictions ($tp+tn+fp+fn$). Accuracy considers all prediction mistakes equally, independent of the specific type of mistake. Other measures show a class-centric view: Precision is the ratio of correctly predicted positive examples and all examples predicted as positive ($\frac{tp}{tp+fp}$), i.e., comparing the number of examples that were correctly assigned to the positive class to the total number of examples that were initially assigned to the positive class. Recall compares the number of correctly predicted positive examples to the number of positive examples, including those incorrectly predicted as negative ($\frac{tp}{tp+fn}$). In medical applications, sensitivity and specificity are commonly used, where sensitivity and recall are equivalent, and specificity is defined as the ratio of examples correctly predicted as negative and all examples predicted as negative ($\frac{tn}{tn+fp}$). Precision, recall/sensitivity, and specificity assess the effectiveness of the algorithm with respect to a single class~\cite{sokolova2006beyond}. The F1-score is the harmonic mean of precision and recall~\cite{sokolova2006beyond}. All six measures produce values in the range of 0 to 1.

The ROC (receiver operating characteristic) curve~\cite{Bradley1997AUC} is a graph showing the classification performance when varying the class discrimination threshold, i.e., the threshold on the continuous classification score above which the positive class is assigned. The ROC plot shows the relation of the true positive rate (or sensitivity) and false positive rate (1 - specificity). 
A random classifier, i.e., a classifier that randomly guesses the label, lies on the diagonal in ROC space. A classifier that is better than random shows a ROC curve above the diagonal. AUC takes the area under the ROC curve as a performance measure. Its value ranges between 0 and 1, with values $> 0.5$ indicating that the classifier performs better than random guessing.

\subsection{Cost-Sensitive Learning}
\label{ssec:rel_costs}
In the performance measures discussed in the previous section, prediction mistakes of the same class have the same influence on the final score. That is, it does not matter on which particular example within a class the prediction was incorrect. However, these measures are oblivious to other characteristics, such as what consequences follow from the individual mistake or how plausible a class label appears to a human.

Cost-sensitive learning acknowledges that different prediction mistakes cause different consequences~\cite{elkan2001cost, kukar1998cost}. For example, in medical diagnosis, falsely assuming a patient has cancer and consequently requesting further tests is a less severe mistake than falsely assuming a patient is healthy and consequently discontinuing further treatment. Each type of prediction outcome ($tp$, $tn$, $fp$, $fn$) receives a ``cost'' that will be added to a grand total during the evaluation process. Instead of optimizing for one of the performance measures described in the previous section, cost-sensitive systems will try to reduce the sum of all costs associated with the prediction outcomes. Cost-sensitive learning is applicable especially in areas where different prediction mistakes severely impact a personal or societal level, including applications in medicine~\cite{wang2018predicting}, hiring~\cite{Yang2017combining}, or credit evaluation~\cite{xiao2020cost}.

Besides penalizing types of prediction outcomes ($tp$, $tn$, $fp$, $fn$), other cost distributions are possible. Turney~\cite{turney2002types} presents a comprehensive overview of cost types that could be included in the training process of machine learning systems. Amongst others, Turney differentiates four types of conditional costs of prediction mistakes, with one being individual costs per example~\cite{turney2002types}, such as the amount of money lost per ignored fraud case in fraud detection. 

\subsection{Ground Truth Labels and Label Noise}
\label{ssec:rel_truth}
For classification tasks, such as topic prediction, medical diagnosis, or fraud detection, a labeled dataset is needed to train and test supervised machine learning classifiers. Each data point in a labeled dataset is assigned to at least one target label, which constitutes the ground truth. The ground truth may be collected together with the data points, e.g., reviews with star ratings~\cite{mcauley2015amazon}, or retrospectively annotated, e.g., objects in a picture~\cite{deng2009ImageNet}. The annotation of a dataset is a complex task fulfilled by human annotators~\cite{Muller2021} who do not always agree on the target label. The disagreement in the labeling process can have various reasons. Annotators' interpretation of a label concept can evolve throughout the annotation process~\cite{kulesza2014structured} or annotators may be inattentive -- not paying close attention or accidentally selecting the wrong label~\cite{Resnick2021}. Other reasons for disagreement could be ambiguity in the task instructions, in the label interpretation, or in the data itself~\cite{Dumitrache2018}. Due to the ambiguity, some data points are difficult to label~\cite{Dumitrache2018, Tsipras2020contextualizing}, leading to uncertainty of the annotators. 

The most common approach to dealing with disagreement between annotators for determining the final ground truth label is the majority vote for a single data point~\cite{Sheng2008}. 
In cases of disagreement, collecting additional annotations for a data point does not necessarily lead to a more reliable majority vote. Although Gurari and Graumann~\cite{gurari2017crowdverge} showed that it is possible to predict annotator agreement and collect more annotations for data points with high disagreement probability, Pavlick et al.~\cite{Pavlick2019} found that despite having a high number of annotations for a single data point, the disagreement ratio usually remains constant. However, regardless of how much disagreement was recorded, the final ground truth often collapses all annotations into a single label~\cite{Dumitrache2018}. This aggregation creates ``label noise''~\cite{algan2021image, Sheng2008, Zhu2004class}, i.e., unreliable ground truth data, on which machine learning systems are trained.

There is extensive research to address the label noise problem from a system perspective (see~\cite{algan2021image, Frenay2014} for an overview). Recent research focuses on considering the full label distribution to reflect human perceptual uncertainty (e.g., in~\cite{Peterson_2019_ICCV, Wang2019}). Gordon et al.~\cite{Gordon2021} introduce their disagreement deconvolution approach to align machine learning classification metrics more closely with user-centric performance measures. Considering the disagreement in annotators' opinions, especially in social computing tasks, they suggest comparing classifier predictions to individual, stable opinions from each annotator. Their algorithm allows computing disagreement-adjusted versions of any standard classification measure, taking each annotators' primary label (stable opinion) as individual ground truth values. Gordon et al. evaluated their algorithm with three social computing tasks and two classic machine learning tasks. Especially for the latter, their approach drastically reduced reported performance (as calculated by classical performance measures). Resnick et al.~\cite{Resnick2021} follow a similar approach: They consider individual human annotators rather than a majority vote of all annotators and introduce a new variable that describes ``how many labels would be needed to get the same expected score that the classifier got''~\cite{Resnick2021}. They suggest using this score to determine whether a system performs well enough.

\subsection{Human Perception of Accuracy and Prediction Mistakes}
\label{ssec:rel_perception_mistakes}
Research found that low perceived accuracy might affect user experience and, especially, trust in such systems~\cite{Dzindolet2002}. Yin et al.~\cite{Yin2019understanding} found significant effects of stated accuracy on people's trust in the system. In their experiment, participants reported more trust in systems that initially claimed to have a high level of accuracy. But this priming effect decreased when users experienced a lower accuracy in practice. Other research found that people stop trusting an algorithm after observing prediction mistakes, even when the algorithm overall outperforms human predictions~\cite{Dzindolet2002, dietvorst2015algorithm}. This might be explained by the fact that when making judgments under uncertainty, individuals often rely on mental short-cuts or rules of thumb (i.e., heuristics) rather than undertaking a thorough and rational analysis~\cite{tversky1974judgment}. One such heuristic is the availability heuristic~\cite{Tversky1973availability}, in which individuals evaluate the probability of cases or events based on their mental availability, i.e., ``by the ease with which relevant instances come to mind''~\cite[p.~207]{Tversky1973availability}. Thus, if multiple cases or events are disproportionally available, the availability heuristic can lead to biased probability judgments (i.e., over- or underestimation), also known as availability bias~\cite{wang2019designing}. Underestimating a classifier's performance does not only reduce trust~\cite{Yin2019understanding} but can also result in users ignoring a system recommendation~\cite{Dietvorst2018}. Nourani et al.~\cite{Nourani2019effects} added explanations to their machine learning system and researched their effect on user trust and perceived accuracy. Different types of explanations had different effects on how participants perceived the system's accuracy, showing that multiple factors can influence users' perception. Roy et al.~\cite{Roy2019} stated that if users have a chance to control the system, the self-reported satisfaction remained constant even when system accuracy is relatively low. 

Investigating single predictions rather than the aggregated view of a classifier's performance, Kocielnik et al.~\cite{Kocielnik2019accept} tested whether a high-recall system (avoiding $fn$) is more accepted than a high-precision system (avoiding $fp$). In their use case, a machine learning system automatically extracted and scheduled meeting appointments from emails. Although other researchers have argued (but not shown) that, in general, false positives ($fp$) are less accepted by users than false negatives ($fn$)~\cite{Kay2015how,Li2008learning}, Kocielnik et al. found that for their application, avoiding false negatives was more important to the users. The perceived accuracy of their system was lower for the high-precision system than for the high-recall system. Even though no general rule can be established, it shows that users weigh different outcomes differently, depending on the associated consequences. Although not providing empirical evidence, Lipton ~\cite{lipton2018mythos} argues that users might be inclined to accept systems that show behavior similar to theirs, i.e., systems which make mistakes only on data points that are difficult to classify for humans. Interestingly, in a clinical text labeling scenario, Levy et al.~\cite{Levy2021} found that domain experts notice if system label recommendations are inadequate but are still likely to accept them for lack of suitable alternatives. In an annotation setting, Tsipras et al.~\cite{Tsipras2020contextualizing} also concentrate on how single classifier predictions are perceived by users. By asking annotators how reasonable they think a prediction is, they compared the performance of different systems on the ImageNet dataset~\cite{deng2009ImageNet}. In their experiment, participants decided whether they agree with predicted labels for multi-object images. Their results showed that non-expert annotators often judged dataset labels as valid even when the prediction was incorrect, which means that the perceived accuracy of the systems was higher than the calculated accuracy. They conclude that just focusing on measuring accuracy is not sufficient to understand system performance in their case on the object recognition task. 

\subsection{Human-Centered Evaluation of Machine Learning Systems}
\label{ssec:rel_perception}
While machine learning measures mainly focus on system performance, human-centered machine learning (HCML) investigates how the users are affected by such systems~\cite{Kaluarachchi2021}. Accordingly, research in HCML focuses on users' evaluations of machine learning systems, including aspects such as users' reliance~\cite{Lu2021} on the system, or trust~\cite{Papenmeier2019, Yin2019understanding} in the system, the perceived explainability~\cite{Smith-Renner2020}, interpretability~\cite{Poursabzi2021}, or  fairness~\cite{Madaio2020, Grgic-Hlaca2018} of the system, and the experienced or perceived accuracy of the system~\cite{Kay2015how, Fairclough2015accuracy, Gordon2021, Tsipras2020contextualizing}. In a recent HCML survey paper, Kaluarachchi et al.~\cite{Kaluarachchi2021} provide a comprehensive overview of user studies with machine learning systems. To investigate machine learning applications from a human-centered perspective, standard HCI methods such as observations, interviews, and questionnaires are used or have been adapted to machine learning scenarios. For example, Gero et al.~\cite{Gero2020} used the think-aloud method and questionnaires to investigate participants' mental models of an AI agent. Other researchers have employed observation studies and interviews to investigate user experience in machine learning settings \cite{Beede2020, Xu2020}. While in-person user studies in labs (like in \cite{Yang2020}) are rather rare, crowdsourcing tasks are commonly used to collect feedback from the users. Common tasks include labeling text or image data (e.g., in \cite{Lu2021}) or assessing given classifications (e.g., in~\cite{Levy2021, Papenmeier2019, Santhanam2020, Smith-Renner2020}). Often, the classification accuracy is systematically varied to avoid unpredictable behavior by the system~\cite{Lu2021, Papenmeier2019, Yang2020}.

In the context of perception of automated systems, Kay et al.~\cite{Kay2015how} present a survey instrument that helps to predict how acceptable users will find the accuracy of a system. In their user study, they applied an adapted version of the extended Technology Acceptance Model questionnaire~\cite{Venkatesh2000theoretical} to assess their newly introduced measure ``acceptability of accuracy''. They argue that, compared with the popular F1-score, their approach correlates more with the user's real-life acceptance of a system's accuracy. While the ``acceptability of accuracy'' is an important addition in the toolbox of HCML evaluation metrics, it does not explain what caused the acceptance or rejection of a system.

\section{Method}
\label{sec:method}
The major goal of the present work is to investigate the role of different types of prediction mistakes on human's perceived accuracy of machine learning systems. Our research is driven by the following research question:
\begin{itemize}
    \item [\textbf{RQ}] Do different types of prediction mistakes of a classifier equally impact users' perceived accuracy of that classifier?
\end{itemize}
The literature has shown that creating ground truth labels for training and evaluation of machine learning systems is challenging (e.g., due to ambiguity)~\cite{Dumitrache2018,Resnick2021,Tsipras2020contextualizing}. Consequently, not all predictions that are deemed incorrect by the ground truth are perceived to be incorrect by users~\cite{Tsipras2020contextualizing}. It remains unclear to what extent different levels of ambiguity of individual data points impact the users' perception of a classifier's accuracy as a whole. We therefore focused on three different levels of data point ambiguity from the user's point of view: data points that are easy, difficult, and impossible to classify for users.

We expected that the more difficult it is for users to identify a data point's label (classification difficulty), the fewer prediction mistakes made by a system are noticed by a user. As a consequence, the viewer prediction mistakes are noticed by a user, the higher is the perceived accuracy of the system. To be more precise, we hypothesized that prediction mistakes on easy-to-classify data points lead to a significantly lower perceived accuracy than prediction mistakes on difficult-to-classify data points [\textbf{H1}]. We also expected that prediction mistakes on difficult-to-classify data points lead to a significantly lower perceived accuracy than those on impossible-to-classify data points [\textbf{H2}]. Consequently, if the same number of prediction mistakes leads to a significantly different perceived accuracy for at least one level of classification difficulty (easy, difficult, impossible), the perceived accuracy should also differ significantly from the standard measure of accuracy (calculated as the number of correct predictions divided by the number of all data points) for at least one level of classification difficulty [\textbf{H3}]. If human annotators disagree on the label of a data point, even correct predictions on impossible-to-classify data points could be perceived as prediction mistakes. We therefore hypothesized that a classifier with 100\% correct predictions will have a perceived accuracy of significantly lower than 100\% if it contains some impossible-to-classify data points [\textbf{H4}].

To test our hypotheses, we first collected information on the classification difficulty for a set of data points in a crowdsourcing task (see Section~\ref{ssec:materials}~``\nameref{ssec:materials}''). Based on the results, we curated two datasets to use them with our five fictive classifiers (see Section~\ref{ssec:conditions}~``\nameref{ssec:conditions}''). In an experimental online user study, we evaluated how participants perceive the accuracy of the five classifiers (see Sections~\ref{ssec:procedure}~``\nameref{ssec:procedure}~-~\ref{ssec:measures}~``\nameref{ssec:measures}''). The experiment had received clearance by the ethics board of our institution. All materials and questionnaires as well as the briefing and debriefing text are available online\footnote{\url{https://git.gesis.org/papenmaa/chi22_perceivedaccuracy}}.

\begin{table*}[b]
\addtolength{\tabcolsep}{2pt}
  \caption{Experimental conditions of the user study.}
  \Description{Table describing the five experimental conditions of the user study, starting with the three conditions with 73\% accuracy, followed by the two conditions with 100\% accuracy.}
  \label{tab:conditions}
  \begin{tabular}[t]{+l^l^l^l^c}
    \toprule\tabhead
    \multirow{2}{*}{Condition} & \multirow{2}{*}{Dataset} & \multirow{2}{*}{Sentence Composition}   & \multirow{2}{*}{Mistakes}  &  Calculated\\
    & & & & \textbf{Accuracy} \\\otoprule
    (A)  \No       & mixed  & 10 easy, 10 difficult, 10 impossible & 8, on easy & 0.73\\
    (B)  \Medium   & mixed  & 10 easy, 10 difficult, 10 impossible  & 8, on difficult & 0.73\\
    (C)  \High     & mixed  & 10 easy, 10 difficult, 10 impossible & 8, on impossible & 0.73 \\\midrule
    (D1) \Cmixed   & mixed & 10 easy, 10 difficult, 10 impossible & 0 & 1.0 \\  
    (D2) \Cclear   & clear & 30 easy         & 0 & 1.0\\\bottomrule 
\end{tabular}
\end{table*}

\subsection{Task and Materials}
\label{ssec:materials}
For experimenting with data points at different levels of classification difficulty, we needed a dataset with a proper (i.e., objectively correct) ground truth that we can enrich with information about the classification difficulty of data points. We, therefore, chose the VACOS\_NLQ dataset~\cite{papenmeier2021dataset} that provides 3,560 product descriptions of laptops and jackets collected from English native speakers. During the collection of the VACOS\_NLQ dataset, each participant was prompted to describe a jacket and a laptop. The ground truth is therefore given by design, not retrospectively annotated, resulting in a proper ground truth even if the texts themselves are ambiguous. Another advantage of the VACOS\_NLQ dataset is that the classification task (assigning product categories to product description) is binary and simple, which allowed us to recruit participants without expert knowledge. The concepts of a ``laptop'' and a ``jacket'' are well-known and easily distinguishable for laypersons. Moreover, laptops and jackets both have unique characteristics (e.g., hood, zipper, technical hardware) but also share some common characteristics (e.g., price, color), yielding the potential for clear sentences as well as ambiguous sentences. 

The product descriptions of the VACOS\_NLQ dataset contain on average 3.7 sentences, which gives enough context to clearly identify the product in most cases. To obtain data points at various levels of classification difficulty, we split the descriptions into single sentences and manually pre-selected 129 sentences (62 jacket sentences, 67 laptop sentences). All pre-selected sentences mention one or two product characteristics and have a length similar to the mean length of all sentences in the dataset (M~=~13.59 words, SD~=~5.98). With the consent of the VACOS\_NLQ dataset authors, the pre-selected sentences and the selection criteria are available online\footnote{\url{https://git.gesis.org/papenmaa/chi22_perceivedaccuracy}}. On the platform Prolific\footnote{\url{www.prolific.co}}, we recruited 54 crowd workers (36~female~(f), 18~male~(m), 0~diverse~(d)) with a mean age of M~=~34.70 years (SD~=~12.85) to classify sentences as either describing a jacket or describing a laptop. Each crowd worker labeled between 20 and 30 sentences. The sentences were drawn at random from the pre-selected pool of 129 sentences. Each sentence was classified by at least 10 workers. Additionally, two attention checks were included in the classification task. All workers passed the attention checks and subsequently received financial compensation of 1.20~GBP for an average completion time of 10 minutes. 

We created three sets of sentences that span the whole range of classification difficulty: 

\begin{itemize}
    \item \textbf{Set 1: easy sentences.} Sentences that are easy to classify for a human, i.e., sentences for which 100\% of the annotators in the crowdsourcing task decided for the correct label (as defined in the ground truth). Example: 
    \begin{quote}
        \textit{``it needs to have lots of pockets and a hood too''} (ground truth label: jacket).
    \end{quote}
    Of the 129 pre-selected sentences, 35 fall in this category (16 jacket sentences, 19 laptop sentences).
    \item \textbf{Set 2: difficult sentences.} Sentences that are difficult to classify for a human but one class is more likely than the other, i.e., sentences for which around 70-80\% of the crowd workers decided for the correct label. Example: 
    \begin{quote}
        \textit{``easy to care for, that is, to clean and to store''} (ground truth label: jacket).
    \end{quote}
    We identified 11 sentences for this set (6 jacket sentences, 5 laptop sentences).
    \item \textbf{Set 3: impossible sentences.} Sentences that are impossible to classify for a human because both classes are equally likely, i.e., sentences for which around 40-60\% of the crowd workers decided for the correct label. Example: 
    \begin{quote}
        \textit{``I am going to buy a new cheap one''} (ground truth label: laptop).
    \end{quote}
    The crowdsourcing task revealed 13 suitable sentences for this set (4 jacket sentences, 9 laptop sentences).
\end{itemize}

A one-way ANOVA confirmed that the sets have a sufficient difference (p~<~.001) in average classification accuracy of annotators (i.e., how many annotators assigned the correct label). Post-hoc tests using two-sided Mann-Whitney U-tests showed that all sets are significantly different from each other (p~<~.001 for all three comparisons). Importantly, the one-way ANOVA did not show a significant difference between sets for number of characters (p~=~.780), number of words (p~=~.780), or lexical density\footnote{Defined as the number of content words (verbs, nouns, adjectives, adverbs) divided by the total number of words in a sentence~\cite{gibson1993towards}.} (p~=~.709). We, therefore, conclude that the three sets differ sufficiently in their classification difficulty for humans, which is, however, not provoked by a difference in lexical form or readability.

\subsection{Experimental Conditions}
\label{ssec:conditions}

To evaluate the effect of prediction mistakes for different sentence types, we constructed two datasets from the sentence sets. The \textbf{clear dataset} consists of 30 sentences taken only from the set of easy cases. The \textbf{mixed dataset} contains a mixture of sentences, with 10 easy sentences, 10 difficult sentences, and 10 impossible sentences. We manually created the output of five fictive classifiers (independent variable) to systematically inject prediction mistakes. We create (A) a system that makes 8 prediction mistakes on the 10 easy-to-classify sentences, (B) a system misclassifying 8 out of 10 difficult-to-classify sentences, and (C) a third system that misclassifies 8 out of 10 impossible-to-classify sentences on the mixed dataset. All three fictive systems (A, B, C) misclassify 8 out of 30 sentences in total and therefore have the same calculated accuracy of 73\%. The prediction mistakes were made equally on both classes, hence four on sentences describing laptops and four on sentences describing jackets. Additionally, we introduced a perfect system that makes no mistakes, yielding a calculated accuracy of 100\% (D1) on the mixed dataset as well as (D2) on the clear dataset. Table~\ref{tab:conditions} shows the resulting five experimental conditions. We decided against training a machine learning system because we wanted to inject structured mistakes, ensuring that all mistakes are made on the same sentence set, e.g., only on the simple-to-classify sentences for condition A. Real-life machine learning systems are likely to make mistakes across all three sentence sets, which would make it difficult to investigate the differences between the sentence sets in our experiment.

\begin{figure*}[t]
    \begin{minipage}{.49\textwidth}
        \centering
        \includegraphics[width=\linewidth]{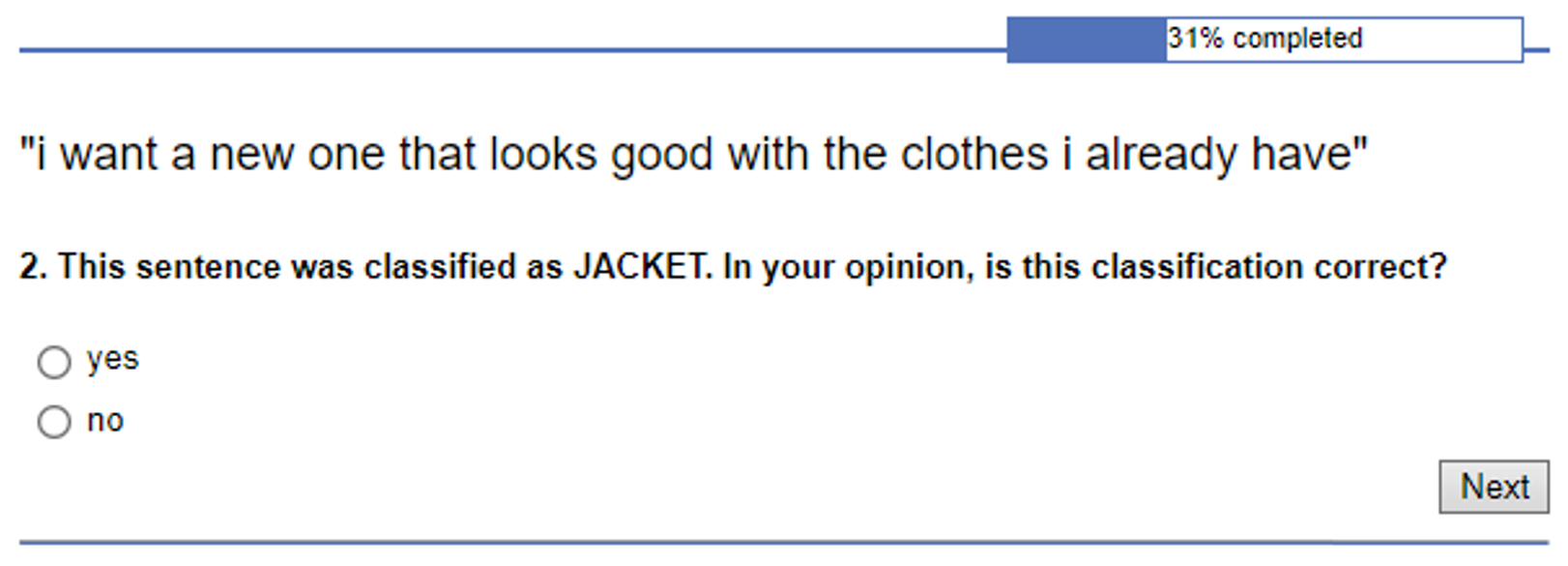}
        \Description{Screenshot of the user study task, showing the easy sentence "I want a new one that looks good with the clothes I already have". Below, the instruction reads "This sentence was classified as jacket. In your opinion, is this classification correct?" with the two options "yes" or "no" for participants to answer.}
        \caption{An easy-to-classify sentence with a correct prediction from the fictive classifier.}
        \label{fig:example_jacket_clear}
    \end{minipage}
    \hfill
    \centering
    \begin{minipage}{.48\textwidth}
        \centering
        \includegraphics[width=\linewidth]{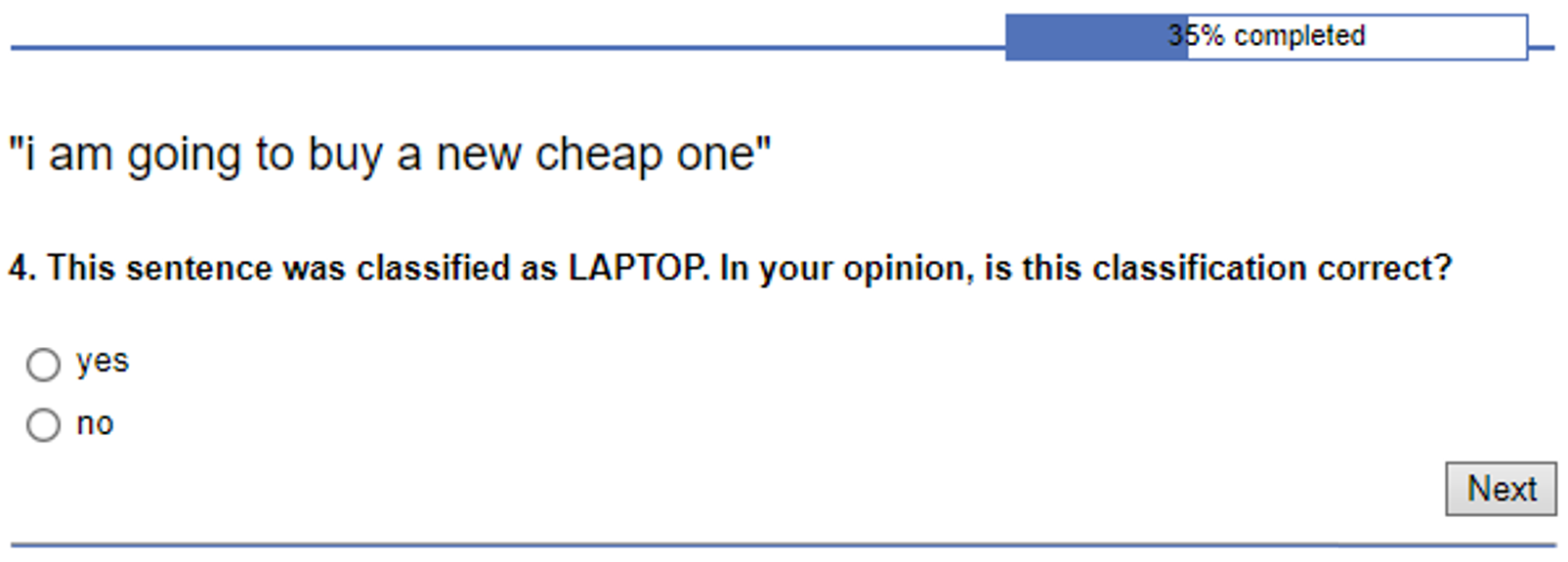}
        \Description{Screenshot of the user study task, showing the impossible-to-classify sentence "I am going to buy a new cheap one". Below, the instruction reads "This sentence was classified as laptop. In your opinion, is this classification correct?" with the two options "yes" or "no" for participants to answer.}
        \caption{An impossible-to-classify sentence with an incorrect prediction from the fictive classifier.}
        \label{fig:example_laptop_unclear}
    \end{minipage}%
\end{figure*}

\subsection{Procedure}
\label{ssec:procedure}
We set up an online user study. After giving informed consent, participants received the task description:
\begin{quote}
    \textit{In this experiment, you will read 30 sentences from product descriptions. The sentences were written by users who were asked to describe a laptop or a jacket they want to buy online. Those sentences have been classified into ``jacket descriptions'' or ``laptop descriptions''. Your task is to assess whether the classification is correct.}
\end{quote}
Participants did not receive information about who or what made the class predictions to avoid priming effects from participants' expectations about a classification system or about an expert. Figures~\ref{fig:example_jacket_clear}~and~\ref{fig:example_laptop_unclear} show examples of the participants' task in the user study. A single task consists of a sentence from the dataset, the fictive classifier's prediction, and a question asking the participants to indicate their agreement or disagreement with the prediction. We chose an interactive task to (1) measure the participants' agreement with the predictions on sentence level, (2) ensure they carefully read the sentences and predictions, and (3) enforce them to form an opinion about the sentence's class. 

All participants passed a training phase in which they assessed the predictions of four sentences: two laptop sentences and two jacket sentences. One sentence of each class was easy to classify and the other one difficult to classify. Afterward, they were informed that they have passed the training and would now start with the main tasks. In the main phase, participants completed 30 tasks with two additional attention checks. The order of sentences was randomized in all conditions to prevent any order effects. Participants in conditions A, B, C, and D1 all interacted with the same 30 sentences from the mixed dataset, whereas participants in condition D2 all saw the same 30 sentences from the clear dataset. Participants did not receive feedback on their classification performance. Following the main phase, participants reported their perceived accuracy, subjective rating of the classifier's performance, perceived agreement, and answered three open questions (see Section~\ref{ssec:measures}~``\nameref{ssec:measures}''). The participants also answered questions about their demographic background (age, gender, and domain knowledge of laptops and of jackets on a 5-point scale). Finally, the participants were debriefed.

\begin{table*}[b]
\parbox{.45\linewidth}{
    \centering
    \caption{Means and standard deviations of the subjective ratings of all classifiers with 73\% calculated accuracy. Comparison of conditions using two-sided Mann-Whitney U-tests with Bonferroni correction. The asterisk (*) denotes significance at $\alpha$ = .05.}
  \label{tab:results_subjective_rating_mixed}
  \Description{This table reports the means and standard deviations of the subjective ratings for condition A, B, and C. We also show the comparison of subjective ratings between conditions in terms of the p-value.}
  \begin{tabular}{+l^l^S^S^S}
    \toprule\tabhead
    \multirow{2}{*}{\bf Condition}   & Subjective & \multicolumn{3}{l}{\bf Comparison} \\
                & {\bf Rating}  & \multicolumn{3}{l}{\bf p-values} \\
    \otoprule
    (A) \No     & 5.29 (1.15) &    ~n/a   &   .449   &   .008* \\ 
    (B) \Medium & 5.49 (0.88) &   .449   &    ~n/a   &   .026* \\ 
    (C) \High   & 5.87 (1.05) &   .008* &   .026* &    ~n/a   \\ 
    \bottomrule
  \end{tabular}
}
\hfill
\parbox{.45\linewidth}{
    \caption{Means and standard deviations of the perceived accuracy of all classifiers with 73\% calculated accuracy. Comparison of conditions using two-sided Mann-Whitney U-tests with Bonferroni correction. The asterisk (*) denotes significance at $\alpha$ = .05.}
  \label{tab:results_perceived_accuracy_mixed}
  \Description{This table reports the means and standard deviations of the perceived accuracy for condition A, B, and C. We also show the comparison of perceived accuracy between conditions in terms of the p-value.}
  \begin{tabular}{+l^l^r^r^r}
    \toprule\tabhead
    \multirow{2}{*}{\bf Condition}   & Perceived & \multicolumn{3}{l}{\bf Comparison} \\
                & {\bf Accuracy}  & \multicolumn{3}{l}{\bf p-values} \\
    \otoprule
    (A) \No     & 0.66 (0.15) & ~n/a\;  & .008* & < .001* \\ 
    (B) \Medium & 0.73 (0.13) & .008* & ~n/a\;   &   .030* \\ 
    (C) \High   & 0.78 (0.14) & < .001* & .030* &    ~n/a\;   \\ 
    \bottomrule
  \end{tabular}
}
\end{table*}

\subsection{Measures}
\label{ssec:measures}
We measured the following dependent variables in the user study:
\begin{enumerate}
    \item \textbf{Calculated accuracy}: accuracy of the fictive classifier with respect to the ground truth. Measured as: \[\frac{\text{number of correct predictions by classifier}}{\text{total number of data points}}\]
    \item \textbf{Human accuracy}: accuracy of the participants with respect to the ground truth. Measured as: \[\frac{\text{number of correct decisions by participant}}{\text{total number of data points}}\] where the number of correct decisions is identified by observing agreement (answering ``yes'', see Figure~\ref{fig:example_jacket_clear}) with correct classifier predictions and disagreement (answering ``no'') with incorrect classifier predictions.
    \item \textbf{Perceived accuracy}: participants' perception of the fictive classifier's accuracy. Collected after interacting with the fictive classifier by asking ``What do you think: In how many cases was the correct product category (LAPTOP / JACKET) displayed?''. Measured as: \[\frac{\text{participants' perceived number of correct classifications}}{\text{total number of data points}}\]
    \item \textbf{Calculated agreement}: agreement of participants with the fictive classifier. Measured as: \[\frac{\text{number of agreements (``yes'' responses)}}{\text{total number of data points}}\]
    \item \textbf{Subjective rating}: participant's assessment of the fictive classifier's performance on a 7-point scale (1~=~``very poor'' to 7~=~``excellent'')
\end{enumerate}
Additionally, after interacting with the systems, we used open questions to ask participants to reflect on their decision-making and perception:
\begin{itemize}
    \item[Q1] ``In which cases did you have difficulties with your decision?'',
    \item[Q2] ``In which cases was it easy to come to a decision?'', 
    \item[Q3] ``Why do you think those sentences were incorrectly classified?'' (Only if the participant indicated that the classification system made at least one prediction mistake)
\end{itemize}

\subsection{Participants}
\label{ssec:participants}
For our user study, we recruited 235 participants via Prolific\footnote{\url{www.prolific.co}}. Only residents of the US, UK, and Ireland being English native speakers were allowed to participate in the experiment. Additionally, we used the Prolific's prescreening options to exclude participants with literacy difficulties. After the experiment, we excluded 10 participants from our data due to low-effort responses, e.g., gibberish responses to the open questions or failed attention checks. Ultimately, we keep N~=~225 valid responses. Participants (171~f, 53~m, 1~d) were randomly assigned to one of the five conditions (N~=~45 in each condition). On average, participants were M~=~35.70~years old (SD~=~13.53~years). Their average domain knowledge of jackets (M~=~3.97, SD~=~0.94) and laptops (M~=~3.95, SD~=~0.91) was comparable, measured on a 5-point scale. The one-way ANOVA did not reveal significant differences between groups with respect to age (F(5,220)~=~1.76, p~=~.139), gender (F(5,220)~=~0.97, p~=~.424), or domain knowledge of jackets (F(5,220)~=~2.42, p~=~.050) or laptops (F(5,220)~=~0.62, p~=~.648). We paid all participants a financial compensation of 1.30~GBP for an average completion time of 12 minutes. The compensation was based on the minimum wage of the UK.


\section{Results}
\label{sec:results}
To answer our research question, we report differences of perceived accuracy between conditions ([\textbf{H1}]), explore how participants perceived the difficulty of the classification task, compare the perceived accuracy with the calculated accuracy ([\textbf{H2, H3}]), and investigate the implications of the dataset composition for the perceived accuracy ([\textbf{H4}]).

We used one-way ANOVAs for single-variable comparisons of conditions (e.g., comparing the perceived accuracy between conditions) with two-sided Mann-Whitney U-tests (with Bonferroni correction) as post-hoc tests. Likewise, when comparing dependent samples, we used Wilcoxon's two-sided signed-rank tests with Bonferroni adjustments. 
All statistical results are interpreted with $\alpha$ = .05 (two-sided).

\begin{table*}[t]
  \caption{The calculated agreement between participants and fictive classifiers per sentence set per condition.}
  \label{tab:results_calculated_agreement}
  \Description{This table reports the means and standard deviations of the calculated agreement scores for conditions A, B, and C. The agreement describes how often participants agreed or disagreed with the predictions during the interaction. We show the agreement score for the whole interaction and for each of the three sentence sets.}
  \begin{tabular}{+l^c^c^c^c}
    \toprule\tabhead
    & \multicolumn{4}{c}{\textbf{Calculated agreement}} \\
    {\bf Condition} & {\bf All} & {\bf Set 1, easy} & {\bf Set 2, difficult}  & {\bf Set 3, impossible}   \\\otoprule
    (A) \No         & 0.64 (0.07) & 0.21 (0.03) & 0.91 (0.10) & 0.80 (0.13) \\
    (B) \Medium     & 0.79 (0.09) & 1.00 (0.01) & 0.58 (0.18) & 0.78 (0.15) \\
    (C) \High       & 0.86 (0.09) & 0.99 (0.03) & 0.90 (0.13) & 0.67 (0.18) \\
    \bottomrule
  \end{tabular}
\end{table*}

\begin{table*}[b]
  \caption{Comparison of traditional measures with perceived accuracy per condition using two-sided Wilcoxon signed-rank tests, asterisk (*) denoting significance at $\alpha$ = .05. Showing means and p-values adjusted with Bonferroni correction.}
  \Description{This table reports the mean of perceived accuracy in all conditions, as well as the calculated scores of standard performance measures such as the accuracy, precision, recall, F1-score, and ROC AUC.}
  \label{tab:results_calculated_measures}
  \begin{tabular}{+l^r@{\hskip 40pt}^r^r^r^r^r^r^r^r^r^r}
    \toprule\tabhead
    \multirow{2}{*}{\bf Condition}   & \multicolumn{1}{l}{\bf Perceived} & \multicolumn{10}{c}{\bf Calculated} \\
                & \multicolumn{1}{l}{{\bf Accuracy}}  & \multicolumn{2}{l}{{\bf Accuracy}} & \multicolumn{2}{l}{{\bf Precision}} & \multicolumn{2}{l}{{\bf Recall}} & \multicolumn{2}{l}{{\bf F1}} & \multicolumn{2}{l}{{\bf ROC AUC}} \\\otoprule
    (A) \No     & 0.66 & 0.73 & <.001* & 0.71 & <.001*   & 0.71 & <.001*  & 0.71 & <.001* & 0.73 & <.001* \\ 
    (B) \Medium & 0.73 & 0.73 & .999\; & 0.71 & .667\;   & 0.71 & .667\;  & 0.71 & .667\;& 0.73 & .667\; \\ 
    (C) \High   & 0.78 & 0.73 & .007* & 0.69 & <.001*   & 0.79 &  .023* & 0.73 & .007* & 0.74 & .007* \\ 
    \midrule
    (D1) \Cmixed & 0.82 & 1.00 & <.001* & 1.00 & <.001* & 1.00 & <.001* & 1.00 & <.001* & 1.00 & <.001* \\
    (D2) \Cclear & 0.95 & 1.00 & <.001* & 1.00 & <.001* & 1.00 & <.001* & 1.00 & <.001* & 1.00 & <.001* \\ 
    \bottomrule
  \end{tabular}
\end{table*}
\subsection{Perceived Accuracy}
\label{ssec:results_perceived_accuracy}
To investigate whether all prediction mistakes are perceived equally by users we asked participants to assess the performance of the classifier subsequent to the 30 interactions. We measured performance in two ways: participants' subjective performance rating on a polarity profile and the perceived accuracy on a numeric scale. 
For the subjective ratings, we provide means and standard deviations as well as the post-hoc test results of comparing the conditions in Table \ref{tab:results_subjective_rating_mixed}. A one-way ANOVA showed a significant difference in subjective ratings (F(5,220) = 3.56, p = .031) between conditions A, B, and C, even though the calculated accuracy was the same in all three conditions (73\%). Participants of condition C rated performance as significantly better than participants in B and A. However, participants in A and B did not significantly differ in their subjective ratings. Concerning perceived accuracy, Table \ref{tab:results_perceived_accuracy_mixed} reports the means and standard deviations for the perceived accuracy per condition. The perceived accuracy (measured between 0 and 30, according to the number of correctly classified instances) provides a more fine-grained view on the systems' performance than the subjective rating (measured between 1 and 7). Similar to the results of the subjective ratings, a one-way ANOVA revealed a significant difference in perceived accuracy (F(5,220) = 93.18, p < .001). The results of the post-hoc test showed a significant difference between all three conditions: The classifier in A that makes obvious prediction mistakes only on easy-to-classify sentences was perceived to be accurate in significantly less instances than the classifiers in B and C. The classifier in C, making prediction mistakes on impossible-to-classify sentences, was perceived to be accurate in significantly more instances than both the classifier in A and B. This shows that all classifiers with 73\% calculated accuracy were perceived to have a significantly different performance, meaning that the type of prediction mistakes had a significant effect on individuals' perceived accuracy of the classifier.

\subsection{Perceived Mistakes}
We collected insights into how participants identified prediction mistakes and how they perceived such mistakes. The answers to the open questions Q1 and Q2 (see Section~\ref{ssec:measures}~``\nameref{ssec:measures}'') give an indication of why sentences in the present dataset were difficult to classify for participants. Participants mentioned three sources of difficulty: (1) Vague formulations (e.g., ``good material'')'. To deal with vagueness, participants developed their own rules throughout the classification task: \textit{``By the end I had decided that I would probably use 'quality' for a jacket''} (P121). (2) Common attributes (e.g., ``brand'' or ``color''): \textit{``on[c]e the word mac was mentioned which could have referred to a mac computer or mac style of jacket''} (P23). (3) Missing context. Some participants commented that the sentences were either too short or not descriptive enough for them to come to a clear decision: \textit{``Where there was no context, it wasn't clear what the sentence was referring to''} (P27). As a consequence of those three sources of difficulties, participants were unsure about the product described in the sentence. However, by the design of our experiment, they were forced to actively agree or disagree with the system's prediction. Participants mentioned that they rather agreed with the system when in doubt: \textit{``Many of the phrases could have been applied to either so I answered affirmatively unless it was obvious that the wording couldn't apply to the category offered''} (P38), and \textit{``There was a few I put yes as just because I felt that it could be either so I would stick with the orig[i]nal classification''} (P121).

To verify this finding, we report the observed agreement of participants with the classifiers per sentence set per condition in Table \ref{tab:results_calculated_agreement}. On the impossible-to-classify sentences (set 3), the classifier in condition C made only two correct predictions. However, the agreement of participants in C on set 3 was 67\%, which is higher than the 21\% observed in A on set 1. A similar trend can be seen in B on set 2, where the classifier likewise made only two correct predictions. Here, participants' average agreement with the classifier was 58\%. As set 2 and 3 contained sentences with higher classification difficulties, participants were more often in doubt about the correct class label and chose to agree with the classifier, leading to a higher agreement with incorrect predictions.

In the open question Q3, participants described why they think the class labels were incorrectly predicted for those sentences. 
Participants stated that the classifier's prediction mistakes can be ascribed to the same difficulties that humans face (see open questions Q1 and Q2), showing that they expect the classifiers to base their decisions on the same grounds as they did themselves. Some participants also described that identifying prediction mistakes was in itself a difficult task: \textit{``It was just my opinion and not definitely incorrect''} (P37). Another participant concluded that the mistake could have been on their side, showing uncertainty about their own decision: \textit{``it could have been for either so there are some i may have got wrong''} (P100). One participant also criticized the binary classification scheme: \textit{``I felt that it was MORE like one over the other''} (P103). These findings suggest that some participants did not perceive the decisions as binary.

\subsection{Traditional Performance Measures}
\label{ssec:results_calculated_accuracy}
Currently, the performance of classification systems is often evaluated against the ground truth using a range of different measures. Table~\ref{tab:results_calculated_measures} displays five performance measures (accuracy, precision, recall, F1, ROC AUC) and compares their calculated score with the perceived accuracy score per condition. For all measures (accuracy, precision, recall, F1, ROC AUC), the calculated score differs significantly from the perceived accuracy in all conditions except condition B (where the perceived accuracy is at 73\%). We conclude that neither calculated accuracy nor any of the examined traditional measures are a suitable representation of the perceived accuracy.

\begin{table*}[b]
\parbox{.45\linewidth}{
  \caption{Means and standard deviations of the perceived accuracy of all classifiers with 100\% calculated accuracy. Comparison of conditions using the two-sided Mann-Whitney U-test and p-values adjusted with Bonferroni correction. The asterisk (*) denotes significance at $\alpha$ = .05.}
  \Description{This table reports the means and standard deviations of the perceived accuracy for condition D1 and D2. We also show the comparison of perceived accuracy between conditions in terms of the p-value.}
  \label{tab:results_perceived_accuracy_both}
  \begin{tabular}{+l^c^r^r^r}
    \toprule\tabhead
    \multirow{2}{*}{\bf Condition}   & Perceived & \multicolumn{2}{l}{\bf Comparison} \\
                & {\bf Accuracy}  & \multicolumn{2}{l}{\bf p-values} \\
    \otoprule
    (D1) \Cmixed     & 0.82 (0.12) & ~n/a\;  & < .001* \\ 
    (D2) \Cclear   & 0.95 (0.10) & < .001* & ~n/a\;   \\ 
    \bottomrule
  \end{tabular}
  }
\hfill
\parbox{.45\linewidth}{
  \caption{Means and standard deviations of the calculated agreement of all classifiers with 100\% calculated accuracy. Comparison of conditions using the two-sided Mann-Whitney U-test and p-values adjusted with Bonferroni correction. The asterisk (*) denotes significance at $\alpha$ = .05.}
  \Description{This table reports the means and standard deviations of the calculated agreement scores for conditions D1 and D2. The agreement describes how often participants agreed or disagreed with the predictions during the interaction. We show the agreement score for the whole interaction and for each of the three sentence sets.}
  \label{tab:results_calculated_agreement_both}
  \begin{tabular}{+l^c^r^r^r}
    \toprule\tabhead
    \multirow{2}{*}{\bf Condition}   & Calculated & \multicolumn{2}{l}{\bf Comparison} \\
                & {\bf Agreement}  & \multicolumn{2}{l}{\bf p-values} \\
    \otoprule
    (D1) \Cmixed & 0.89 (0.08) & ~n/a\;  & < .001* \\ 
    (D2) \Cclear & 0.99 (0.03) & < .001* & ~n/a\;   \\ 
    \bottomrule
  \end{tabular}
  }
\end{table*}

\subsection{Dataset Composition}
\label{ssec:results_dataset_composition}
Participants also interacted with classifiers that made no mistakes. This was the case in condition D1 on the mixed dataset, containing 30 sentences that are easy, difficult, and impossible to classify, and in condition D2 on the clear dataset with 30 easy-to-classify sentences. To test whether the mere presence of difficult- and impossible-to-classify sentences reduces the perceived accuracy, we compared the perceived accuracies reported in D1 and D2. Table \ref{tab:results_perceived_accuracy_both} shows that the perceived accuracy of D2 was significantly higher than the perceived accuracy of D1. We, therefore, conclude that the perceived accuracy is significantly influenced by the dataset composition. The presence of difficult- and impossible-to-classify sentences reduced the perceived accuracy, even if no prediction mistakes were made. However, in D2, the perceived accuracy of 95\% indicates that the classifier was perceived to make 1.5 incorrect predictions on average. It should be noted that during the classification task, participants had to complete two attention checks that looked similar to the sentences. The sentence text stated that the question is an attention check, while the question (usually asking whether they agreed with the class label) instructed participants to answer ``no''. It is possible that answering ``no'' to the attention checks influenced participants' perception, resulting in the perceived average prediction mistake rate of 1.5 for the perfect classifier on the clear dataset. However, the attention check mechanism was constant in conditions, which should equally influence the perceived accuracy in all conditions. 

We also compared the perceived accuracy to how often participants agreed with the classifier (see Table \ref{tab:results_calculated_agreement_both} for the calculated agreement in D1 and D2). While the agreement shows how often participants actively agreed with the classifier's prediction (binary), the perceived accuracy reflects how accurate participants think the classifier was (potentially continuous). Both metrics are measured as $X$ out of 30. In both conditions D1 and D2, the perceived accuracy is significantly lower than the actual agreement (p~<~.001. for both conditions). Furthermore, although there are no prediction mistakes present in either of the two conditions, the calculated agreement of D1 is significantly lower than the one of D2 (p < .001).

\section{Discussion}
The present user study was designed to determine whether different types of prediction mistakes of a classifier equally impact users'  
\vfill\eject\noindent
perceived accuracy of the classifier. In the remainder, we discuss our findings on the perceived accuracy of classifiers that make different types of prediction mistakes. Moreover, we discuss whether existing performance measures are a suitable representation for the accuracy that users believe to have experienced. Finally, we look at the results of investigating the influence of dataset composition on perceived accuracy and place our findings in a broader context of implications for practitioners and the field of human-centered machine learning.

\subsection{Not all prediction mistakes influenced the perceived accuracy equally}
We first investigated how accurate participants believed the system to be. 
In line with our expectations, our results show that even though all three classifiers had the same calculated accuracy, each making eight prediction mistakes on 30 sentences, they brought about significantly different levels of perceived accuracies. Specifically, our results confirm hypothesis [\textbf{H1}], because the perceived accuracy in condition A (where prediction mistakes were made on easy-to-classify sentences) was significantly lower than in conditions B (prediction mistakes on difficult-to-classify sentences) and C (prediction mistakes on impossible-to-classify sentences). Our results also confirm hypothesis [\textbf{H2}], as the perceived accuracy in condition B was significantly lower than in condition C. To understand why not all prediction mistakes influenced the perceived accuracy equally, we asked participants why they think the classifier made mistakes. In their answers, participants reported that their decision is not always binary (``correct'' and ``incorrect''). They perceived the decision to be more nuanced (one is ``more correct'' than the other, but both are ``correct''). In some cases, they doubted their own judgement when confronted with a prediction mistake.
In the open questions, participants also reported to be inclined to agree with the classifier's prediction when in doubt, i.e., when the difficulty of classification for a human increases. The tendency of agreeing when being in doubt also shows in the results of the calculated agreement. For impossible-to-classify sentences, the agreement with the fictive classifier that made eight incorrect and two correct predictions was at 67\%. These findings show that the classifier's (incorrect) predictions impaired participants' decisions for the correct class label, as one would expect to see an agreement of 20\%. This means that in cases were sentences were difficult or impossible to classify, participants tended to agree with the classifier when being in doubt -- even if the predictions were incorrect. A similar observation has been made by Levy et al.~\cite{Levy2021}, who found that participants accepted inappropriate predictions in lack of more suitable alternatives. Interestingly, participants justified their behavior by challenging the definition of ``correct''. They indicated that for some sentences, both class labels were equally likely and should therefore both be considered correct.

\subsection{The calculated measure of accuracy does not accurately reflect perceived accuracy}
In hypotheses \textbf{H1} and \textbf{H2}, we expected and confirmed a significant difference in perceived accuracy between conditions A, B, and C. Since the classifiers in all three conditions make eight prediction mistakes on 30 predictions, hence having a calculated accuracy of 73\%, we hypothesized that the perceived accuracy differs significantly from the calculated accuracy in at least one condition [\textbf{H3}]. Our results indicate that hypothesis [\textbf{H3}] can be accepted, as the calculated accuracy differed significantly from the perceived accuracy in four out of five conditions. Only in condition B (prediction mistakes on difficult-to-classify sentences), no significant difference between calculated and perceived accuracy was shown. We further see in our results that the perceived accuracy in condition A (prediction mistakes on easy-to-classify sentences) was significantly lower (66\%) than the calculated accuracy (73\%). This indicates an underestimation of the classifier's performance. Contrarily, in condition C (prediction mistakes on impossible-to-classify sentences), participants over-estimated the performance of the classifier, with the perceived accuracy (78\%) being significantly higher than the calculated accuracy (73\%). These findings again support the conclusion that the type of prediction mistake (with respect to classification difficulty) influences how users react to a classifier, potentially resulting in under- or overestimation of the actual accuracy. A possible explanation for this observation is the availability bias~\cite{Tversky1973availability, wang2019designing}: Prediction mistakes that were easy to recognize as such (e.g., mistakes on easy-to-classify sentences) are mentally more available than prediction mistakes that were difficult to classify. Clear mistakes might be prevalent when forming an opinion on the classifier's performance, hence having a stronger influence on perceived accuracy.

For other traditional measures used to evaluate the performance of a system (precision, recall, F1, ROC AUC), we also found significant differences between the calculated score and the perceived accuracy in all conditions except condition B. It should be noted that although accounting for some characteristics of a dataset such as class imbalance (F1) or for a type of outcome, such as $fp$ in precision or $fn$ in recall, all those measures treat all prediction mistakes equally. Furthermore, none of these measures considers how a human perceives individual data points, e.g., with respect to their classification difficulty. We therefore conclude that none of the discussed measures is a suitable representation for perceived accuracy.

\subsection{The composition of the dataset itself influences the perceived accuracy}
We hypothesized that even a classifier that does not make any prediction mistakes will have a perceived accuracy that is significantly lower than 100\% if the dataset contains some impossible-to-classify data points [\textbf{H4}]. The statistical analysis of the perceived accuracy in D1 (82\%) and D2 (95\%) reveals a significant difference, which confirms hypothesis [\textbf{H4}]: The mere presence of difficult- or impossible-to-classify sentences reduced the perceived accuracy. Interestingly, participants also thought that the classifiers in both conditions made significantly more mistakes than what they indicated during the interaction (by agreeing or disagreeing with the prediction). This discrepancy might have been caused by sentences where participants were in doubt and therefore had agreed with the prediction, but kept in mind that they were unsure. This uncertainty could have played a role in the retrospective judgement of the classifier's performance.

\subsection{Implications}
The implications of our findings are twofold. First, a metric that accounts for classification difficulty is needed to accurately reflect the users' perception of a classifier's performance. This measure could either be used during evaluation to draw a more accurate picture of how the performance of a classifier is perceived, or during training to optimize a classifier's behavior for perceived accuracy. In both cases, a ``perception weight'' should be assigned to each data point, quantifying the reduction in perceived accuracy that this data point brings about if misclassified. In traditional measures such as accuracy, F1, or ROC AUC, all mistakes have the same weight. The field of cost-sensitive learning~\cite{Thai2010cost} has introduced the concept of cost matrices, that could be leveraged to account for individual weights~\cite{turney2002types} and reduce the gap between system-oriented and user-oriented evaluations~\cite{lipton2018mythos}. A cost-sensitive evaluation metric with individual ``perception weights'' might better reflect how a classifier is perceived by the users. While a new metric would help to understand the user's view on a system, optimizing the system behavior for perceived accuracy should be considered with care, as it holds the potential for manipulating users by fostering overestimation of the classifier's performance, possibly leading to over-reliance, inappropriate user trust, and deception. Previous works have made first steps towards human-centered measures for machine learning systems, but have either focused on an aggregated view rather than on differences between individual data points~\cite{Kay2015how,Resnick2021} or present a theoretical concept that was not tested in user studies~\cite{Gordon2021}. 

Second, our findings on dataset composition indicate that the dataset itself plays an important role in how a user perceives a classifier's performance. If practitioners want to understand how their system will be perceived by users, it is not enough to only focus on how many mistakes a system makes. They need to also consider the dataset composition: The mere presence of classification difficulty reduces the perceived accuracy, potentially leading to a significant underestimation of the performance. As a consequence, users could lose trust in the system's capabilities. Dietvorst, Simmons, and Massey \cite{Dietvorst2018} found that an underestimation can lead to ignorance of the system's suggestions, even in case that the system outperforms human accuracy. In addition to utilizing methods to enhance ground truth robustness (see~\cite{algan2021image,Frenay2014,Peterson_2019_ICCV,Wang2019}), we encourage practitioners to also investigate a dataset's composition concerning the classification difficulties of its data points. It is important for practitioners to understand how a system will be perceived by users when in production, which is why we suggest including data point classification difficulty (from a user's point of view) in the assessment of machine learning systems in practice.

\section{Limitations and Future Work}
The study is subject to several limitations. First, by asking for the agreement or disagreement on each prediction, participants were forced to form an opinion for each sentence. Although this should make it easier for participants to assess the accuracy of the system at the end of the interaction, it might differ from how users would use a decision-support system in the real world. Replicating our study without asking for an explicit agreement or disagreement for each sentence could deliver additional insights into the opinion-forming process of users and the generalizability of our findings.

Second, some participants described the classification decision as being non-binary. In our study, however, participants were forced to give a binary decision (``yes'' or ``no'') rather than indicating their level of agreement. Based on our findings, we suggest that future studies consider using interval scales to indicate the level of agreement rather than a binary decision.

Third, we have used only one use case in our study. To broaden the picture of how classification difficulty impacts the perceived accuracy in various domains, future research should investigate additional textual datasets, e.g., spam detection or news articles. Those use cases offer data points with varying classification difficulties, as well as different type of mistakes: Authors of spam emails continuously find new ways of disguising their messages (e.g., writing ``biitcoiin'' instead of ``bitcoin'') that do not appear in the training data, leading to obvious mistakes of a model on easy-to-classify data points. The transferability of our findings to other data types, e.g., images or audio data, should also be investigated. However, future studies should ensure that the ground truth was, similar to our setup, already gathered during data collection and should avoid datasets that are only retrospectively annotated. 

Fourth, by not disclosing to the users where the predictions stemmed from, we aimed to avoid priming the judgement of our participants. Introducing either a system or a human expert that produces the predictions could result in over-reliance or under-reliance~\cite{Dzindolet2002}. It could be interesting to repeat the experiment with an altered task introduction, explicitly mentioning either a human expert or a machine learning system. 


Finally, our study also revealed that participants tended to agree with the system when in doubt, i.e., when the difficulty of decisions increased. To develop a full picture of the agreement behavior, future studies are needed that explore the influencing factors of decisions taken under uncertainty.

\section{Conclusion}
In this paper, we investigated how the individual classification difficulty of data points influences perceived accuracy of a classifier in a binary classification task. We conducted an online user study with 225 participants who received 30 predictions of one out of five fictive classifiers and collected the perceived accuracy from participants. We investigated three systems with equal accuracy (73\%) that made mistakes on different types of data points (easy-to-classify, difficult-to-classify, impossible-to-classify). Additionally, we compared perfect classification performance at 100\% accuracy on a dataset with only easy-to-classify cases to a dataset with mixed data points. Our findings show that (1) mistakes do not equally impact perceived accuracy (e.g., mistakes on impossible-to-classify data points result in a higher perceived accuracy than mistakes on easy-to-classify data points), (2) the perceived accuracy can significantly differ from calculated accuracy, and (3) the composition of the dataset itself (in terms of the classification difficulty of the data points) has an impact on perceived accuracy. We suggest that predictions on user acceptance of a system should therefore not be based on traditional accuracy metrics such as precision, recall, or the F1-score, as they are not an accurate representation of how users perceive the accuracy of a classification system.

\begin{acks}
This work was partly funded by the DFG, grant no. 388815326; the VACOS project at GESIS.
\end{acks}

\bibliographystyle{ACM-Reference-Format}
\bibliography{chi22percaccbib}

\appendix

\end{document}